\documentclass[journal]{IEEEtran}

\makeatletter
\def\endthebibliography{%
	\def\@noitemerr{\@latex@warning{Empty `thebibliography' environment}}%
	\endlist
}
\makeatother

\usepackage{cite}

\usepackage[pdftex]{graphicx} 

\usepackage[caption=false,font=footnotesize]{subfig}
\usepackage[export]{adjustbox}

\usepackage{amsmath}
\usepackage{amsfonts}
\usepackage{bm}

\usepackage{filecontents}
\usepackage{amssymb}
\usepackage{color}

\usepackage{epsfig}

\usepackage{verbatim}

\usepackage{url}

\usepackage{algorithm, tabularx}

\usepackage{lettrine}

\usepackage{lipsum}

\usepackage{siunitx}

\usepackage{soul,color}

\usepackage{mathrsfs}

\usepackage{array}

\usepackage[inline]{enumitem}

\usepackage{dblfloatfix} 

\usepackage{algorithm}

\usepackage[noend]{algpseudocode}

\makeatletter
\DeclareRobustCommand{\iscircle}{\mathord{\mathpalette\is@circle\relax}}
\newcommand\is@circle[2]{%
  \begingroup
  \sbox\z@{\raisebox{\depth}{$\m@th#1\bigcirc$}}%
  \sbox\tw@{$#1\square$}%
  \resizebox{!}{\ht\tw@}{\usebox{\z@}}%
  \endgroup
}
\makeatother

\usepackage{setspace}
\usepackage{multirow}
\usepackage{array}
\newcolumntype{L}[1]{>{\raggedright\let\newline\\\arraybackslash\hspace{0pt}}m{#1}}
\newcolumntype{C}[1]{>{\centering\let\newline\\\arraybackslash\hspace{0pt}}m{#1}}
\newcolumntype{R}[1]{>{\raggedleft\let\newline\\\arraybackslash\hspace{0pt}}m{#1}}

\begin{document}
	
	\title{Pointing-and-Acquisition~for~Optical~Wireless~in~6G: From Algorithms to Performance Evaluation}

	
	\author{Hyung-Joo Moon,~\IEEEmembership{Graduate Student Member,~IEEE}, Chan-Byoung Chae,~\IEEEmembership{Fellow,~IEEE},\\Kai-Kit Wong,~\IEEEmembership{Fellow,~IEEE}, and Mohamed-Slim Alouini,~\IEEEmembership{Fellow,~IEEE}
	}
	
	{}
	

	\maketitle

	\begin{abstract}

The increasing demand for wireless communication services has led to the development of non-terrestrial networks, which enables various air and space applications. Free-space optical (FSO) communication is considered one of the essential technologies capable of connecting terrestrial and non-terrestrial layers. In this article, we analyze considerations and challenges for FSO communications between gateways and aircraft from a pointing-and-acquisition perspective. Based on the analysis, we first develop a baseline method that utilizes conventional devices and mechanisms. Furthermore, we propose an algorithm that combines angle of arrival (AoA) estimation through supplementary radio frequency (RF) links and beam tracking using retroreflectors. Through extensive simulations, we demonstrate that the proposed method offers superior performance in terms of link acquisition and maintenance.

	\end{abstract}

	\IEEEpeerreviewmaketitle

	\section{Introduction}
	
	As network users demand higher data rates and lower service latency, sixth-generation (6G) networks aim to address various applications and functional nodes with different critical constraints.
	Free-space optical (FSO) communications have emerged as promising solutions for future wireless networks due to their significant advantages, including wide bandwidth, immunity to eavesdropping, long link distances, and no interference with radio-frequency (RF)-based terrestrial networks~\cite{202304commag}. Aerial FSO communications have the potential to become a key technology that integrates terrestrial and non-terrestrial networks (NTNs).

\begin{figure*}[t]
	\begin{center}
		{\includegraphics[width=1.15\columnwidth,keepaspectratio,frame]
			{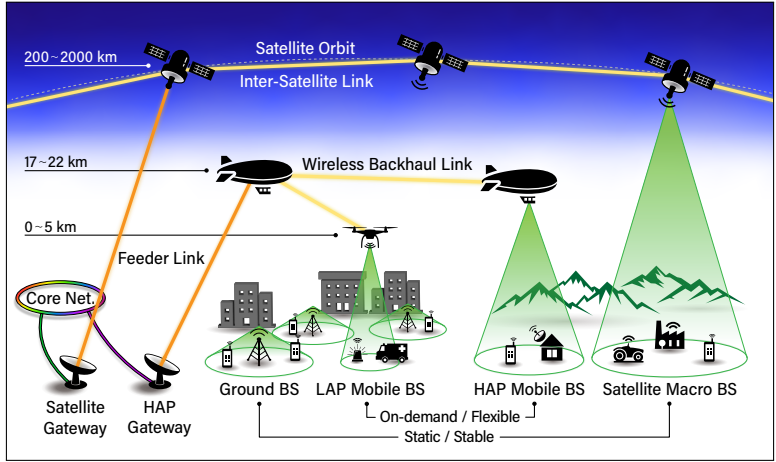}%
			\caption{The role of optical wireless backhaul links in 6G networks is illustrated. Low altitude platforms (LAPs), high altitude platforms (HAPs), and satellites can be supported by FSO links to provide connectivity to unserved areas.}
			\label{fig01}
		}
	\end{center}
	\vspace{-15pt}
\end{figure*}

	
	FSO backhaul link applications can overcome the installation cost and environmental constraints. As depicted in Fig.~\ref{fig01}, the applications include data traffic offloading, coverage extension, and mission-critical services enabled by the flexible deployment of mobile base stations, creating a variety of on-demand cells that constitute heterogeneous networks~\cite{202304taes}. In particular, low-altitude and high-altitude platforms could play a vital role in the 6G network, acting as on-demand agile base stations. While concerns may arise regarding the flexible deployment of the aerial platforms and potential interruptions to other radio resources, FSO communications address these issues, making it a promising enabler for future integrated network design.

\textcolor{black}{The successful implementation of long-distance FSO communications largely depends on the performance of pointing, acquisition, and tracking (PAT) systems~\cite{201804cst}. A PAT system is both fundamental and challenging to ensure the viability of FSO links in 6G networks, where network outages are not permissible or must be anticipated in advance~\cite{202304commag}.
The design of the PAT system has evolved over decades of theoretical and experimental research. However, current technical standards and system-level designs mostly consider satellite communications or static aerial communications~\cite{ccsds}. In order to contribute to the broader applicability of FSO communications within dynamic future wireless networks, we introduce a novel high-level approach for pointing-and-acquisition tailored to near-earth 6G NTNs. To the best of our knowledge, this is the first comprehensive study of design issues in bidirectional FSO communications between ground and air nodes from a PAT perspective.}


Throughout this article, we thoroughly discuss the following topics.

1. We introduce a baseline PAT system for vertical FSO links equipped with conventional detectors and actuators. The mechanisms of link acquisition and link maintenance for bidirectional communications are defined based on clear foundations.

2. We outline various considerations for the PAT of the vertical FSO link based on a survey of experimental works. The external effects of the atmosphere, internal limitations of aircraft payload, and other factors impact the link-acquisition and link-maintenance processes.

3. We propose novel pointing-and-acquisition algorithms for aircraft communications in 6G. The considerations for bidirectional FSO communications are mitigated through the proposed techniques added to the baseline system. The simulation results show that our method surpasses the baseline method in terms of robustness and agility.

	\section{PAT for Vertical Aerial FSO Links}

	\textcolor{black}{There are no specific standardizations for PAT system design, largely because link acquisition and link maintenance predominantly rely on local processing, and the system requirements vary greatly depending on the mission. Nevertheless, a range of experimental studies have effectively demonstrated a variety of promising system architectures. Based on this knowledge, this section presents the baseline PAT system for aerial FSO links, which is reasonably designed using conventional devices and mechanisms to facilitate bidirectional communications. Building on this foundational approach, we further develop a PAT system that ensures fast and robust link connections.}

	\subsection{Baseline of the Pointing, Acquisition, and Tracking System}
	\label{baseline}


	\textcolor{black}{The PAT system consists of open-loop coarse pointing (OLCP), closed-loop coarse pointing (CLCP), and fine tracking~\cite{201804cst}. The OLCP is an initial link acquisition process that relies on the prior positioning information of the terminals. In this stage, open-loop beam control is implemented to locate the other terminal and establish a beacon link connection. Then, the CLCP supports closed-loop link maintenance by utilizing feedback from the receiver image sensor. In order to mitigate pointing disturbances with higher frequencies, the fine-tracking system also employs closed-loop beam control to suppress pointing disturbance over a wider bandwidth than the CLCP. More specifically, we now introduce the baseline PAT algorithm of the ground-air bidirectional link, as depicted in Fig.~\ref{fig02}.}


	\subsubsection{OLCP}
	When a vertical FSO link is scheduled for a gateway and aircraft, the aircraft first transmits its positioning information to the gateway via RF links. The gateway then sets an area where the aircraft can be located and scans the area with a beacon beam. The aircraft receives the beacon beam through a focal plane array (FPA) to estimate the incident direction of the beam. Generally, charge-coupled devices or complementary metal-oxide-semiconductor cameras are used as FPAs to offer a wide field of view (FoV). The aircraft controls the gimbal through this estimation to transmit the beacon laser to the gateway. By detecting the downlink beacon beam, the gateway also controls the gimbal to align the transmit and receive pointing directions.
	
	\subsubsection{Fine Tracking}
\begin{figure*}[t]
	\begin{center}
		{\includegraphics[width=2\columnwidth,keepaspectratio,frame]
			{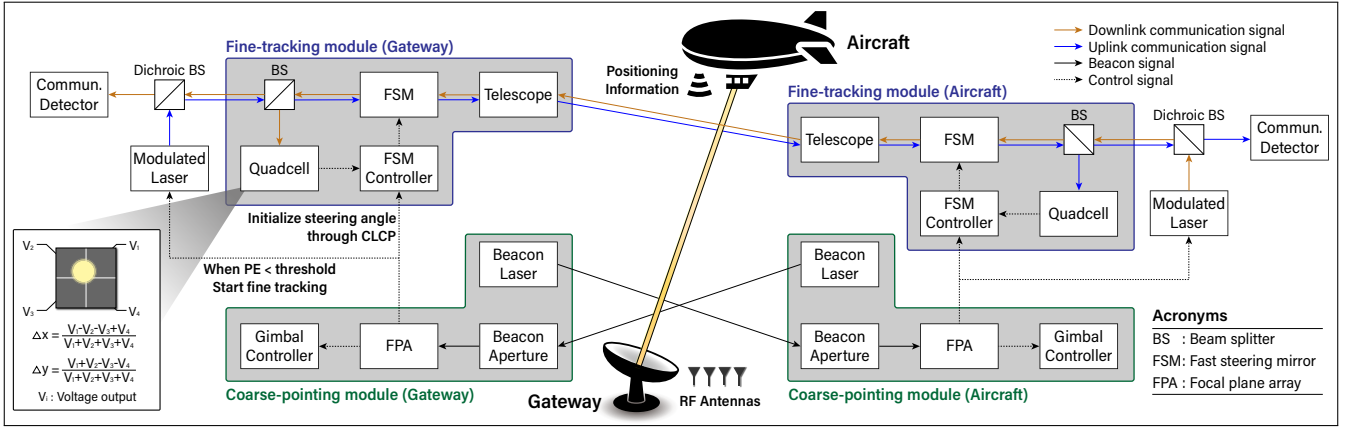}%
			\caption{The baseline PAT system for a vertical and bidirectional FSO link are described.}
			\label{fig02}
		}
	\end{center}
	\vspace{-15pt}
\end{figure*}

	If the beam alignment is accurate enough to be within the FoV of the communication detector, both terminals transmit a communication beam. To maintain the link, a quadcell measures the pointing error of the received communication beam in the horizontal and vertical directions, using the difference in the received power in each quadrant. Using the feedback signal from the quadcell, a fast steering mirror (FSM) controller controls the FSM to compensate for the pointing error of both the transmitting and receiving beams. At the quadcell, detecting communication beams using a beam splitter~\cite{201804cst} and detecting beacon beams~\cite{2020ssc} are both possible. As shown in Fig.~\ref{fig02}, we adopt communication beam detection in the baseline system. This approach helps avoid wavelength-selective effects and extra calibration errors by implementing misalignment and communication detection within the same wavelength and optical system.


	
	\subsubsection{CLCP}
	Along with the fine-tracking process, the CLCP process manages the pointing direction of the entire PAT payload using the gimbals. It maintains the link within the FoV of both the beacon and communication receivers and protects the maximum dynamic range of the FSMs by initializing their tilting angles. The CLCP and fine-tracking processes involve trade-offs between tracking accuracy and dynamic range. Consequently, both closed-loop controls are crucial for enabling bidirectional communications, particularly for low-altitude mobile aircraft that require a broad operating angle range and tracking bandwidth due to the extreme transmission conditions~\cite{2014advancedfso}.

	\subsection{Device Considerations}
	
	Candidates such as modulating retroreflectors (MRRs) and liquid crystals are under research as tracking actuators for cost-efficient PAT payloads~\cite{201804cst}. However, using these cost-efficient devices makes implementing bidirectional links challenging, as specified in Section~\ref{paylim}. Therefore, we adopt conventional spot position detection and mechanical beam steering methods as the baseline PAT system. The cooperative tracking of fine-tracking and CLCP systems compensates for both short- and long-term angular fluctuations caused by mobility, posture change, and mechanical jitters of the aircraft. Stable suppression of pointing error allows higher received signal power through using narrower communication beams and prevents link outages due to various factors during flight.

	\subsection{Hybrid RF/FSO Link}
	
	
	Hybrid RF/FSO communications are one of the most active research areas in optical wireless communications. The integration of these two different systems offers robustness, flexibility, and extensive communication capacity~\cite{2015wcom}. Although many works highlight improvements in communication performance, the additional RF link also provides advantages in the PAT aspect. First, as discussed in Section~\ref{baseline}, the RF link allows the real-time exchange of aircraft positioning information, which is essential for link acquisition~\cite{201804cst,boeing90}. Second, the exchange of network-level control information, such as link switching between gateways or aircraft, can be fully supported by RF links~\cite{202304taes}. This ensures reliable and seamless transfer of control plane data between the nodes, while point-to-point FSO links support the exchange of user data at higher data rates. Lastly, RF links can serve as a supplementary data link parallel to FSO communications~\cite{2015wcom}. Due to these advantages and its necessity to enable flexible link scheduling in aerial networks, our baseline PAT system includes a supplementary RF link.

	\section{PAT Considerations for Aerial FSO Communications}

	\subsection{Atmospheric Effects}
	
	\textcolor{black}{Atmospheric conditions significantly impact signal quality in aerial FSO communications.}
	\textcolor{black}{Among various atmospheric effects, attenuation, beam wander, and scintillation can pose major challenges in PAT systems.}

	\subsubsection{Attenuation}
	\textcolor{black}{During the propagation of a beam, various molecules and small particles absorb and scatter the electromagnetic wave, leading to attenuation. Following the Beer-Lambert law, these atmospheric effects reduce the signal strength exponentially over distance~\cite{2014advancedfso}. Attenuation can vary significantly depending on weather conditions, such as rain, fog, and clear weather, ranging from $0.5$ to over $30$~dB/km. Consequently, the power loss of communication and beacon beams can dramatically increase due to weather changes or cloud movements. In such cases, a network management system should command the aircraft to move to a clearer site or schedule other gateways and aircraft to establish a new routing path~\cite{202304taes}.}
	
	\subsubsection{Beam Wander}

	\textcolor{black}{Beam wander refers to the phenomenon where the beam path shifts due to an eddy larger than the beam size. This random movement of the beacon and communication beam footprints on the receiver plane leads to considerable power loss. Near the Earth surface, the path alteration due to beam wander extend up to hundreds of microradians~\cite{201801access}. However, the impact of beam wander is less critical at high elevations, where the atmospheric density is significantly lower. Furthermore, the authors of~\cite{201801access} showed that the chromatic effect of beam wander is negligible. This demonstrates that the pointing errors of the beacon beam and communication beam are highly correlated, and the cooperation of coarse pointing and fine tracking can ensure the alignment of the two links concurrently.}
	
	
	\subsubsection{Scintillation}
	Scintillation is an intensity fluctuation from small eddies causing random intra-beam phase disturbance. It creates rapid and severe fluctuations in the received signal power and significantly affects the signal quality of both beacon and communication beams. Moreover, it is essential to consider that wider beams are less affected by scintillation when determining the beam divergence angle of FSO communication systems~\cite{1995ao}.

\begin{table*}[t]
	\centering
	\caption{Considerations and Requirements for Bidirectional FSO Communications with Aircraft}
	\footnotesize
	\label{tbl01}
	\begin{tabular}{|p{1.6cm}|p{2.7cm}|p{3.1cm}|p{2.7cm}|p{3.1cm}|p{3.1cm}|}
	\hline
	PAT stage & \multicolumn{2}{p{5.8cm}|}{OLCP} & \multicolumn{2}{p{5.8cm}|}{CLCP} & Fine tracking\\
	\hline
	Platforms & Static platforms & Mobile platforms & Static platforms & Mobile platforms & Both\\
	\hline
	Atmospheric effects & \multicolumn{2}{p{5.8cm}|}{$\bullet$ Channel assessment and link scheduling are required before the OLCP process. $\bullet$ OLCP outage occurs due to high atmospheric loss depending on the weather.}  &  \multicolumn{2}{p{5.8cm}|}{$\bullet$ The CLCP system quickly recovers the fine tracking outage due to atmospheric effects. $\bullet$ When the beacon link is also out, the system begins with the OLCP.} & $\bullet$ The fine-tracking outage can occur during foggy weather due to a power outage.\\
	\hline
	Payload capacity~\cite{201009jstqe} & $\bullet$ Balloons and blimps usually have extensive payload limit to carry devices with enhanced pointing capability. & $\bullet$ Fixed- and rotary-wing aircraft usually have limited payload capacity. $\bullet$ GNSS and beacon-based pointing capability are necessary to respond to the movement. & $\bullet$ Usually capable of robust gimbal-based CLCP with extensive payload. & $\bullet$ To contend with vast flight trajectories, a gimbal-based CLCP or other advanced solutions with large FoV and dynamic range is necessary despite the limited capacity. & $\bullet$ Various payload-saving fine-tracking strategies (using MRR, liquid crystal, or beaconless method) can be considered~\cite{201804cst}.\\
	\hline
	Positioning, mobility, and posture changes & $\bullet$ Faster link acquisition is allowed for static platforms with GNSS capability~\cite{202202wcl}. & $\bullet$ Outdated GNSS information can cause inaccurate beacon beam scanning. $\bullet$ Gateways can consider the velocity of aircraft when determining the scanning area~\cite{2014advancedfso}. & - & $\bullet$ CLCP protects the FoV and dynamic range of the fine tracking system, where the mobility and posture changes cause excessive pointing errors~\cite{boeing90}. & $\bullet$ PAA is negligible in most aerial scenarios~\cite{2020ssc}. $\bullet$ Fine tracking systems are exposed to the risk of exceeding FoV or dynamic range.\\
	\hline
	Others~\cite{2020ssc} & \multicolumn{2}{p{5.8cm}|}{$\bullet$ Open-loop gimbal pointing error, aircraft body pointing error, and attitude sensor misalignment.} & \multicolumn{2}{p{5.8cm}|}{$\bullet$ Closed-loop gimbal pointing error, aircraft body pointing error, FPA measurement error, and misalignments.} & $\bullet$ Quadcell measurement error and misalignment, FSM control error, and residual mechanical jitter.\\
	\hline
	\end{tabular} 
	\vspace{-10pt}
\end{table*}

	\subsection{Payload Limit}
	\label{paylim}
	
	There are large differences in payload capacity depending on aircraft type and size, which significantly constrains available power and equipment. Thus, various PAT techniques utilizing lightweight and cost-effective devices have been proposed to support aircraft with limited payload capacity.

	\subsubsection{Modulating Retroreflectors}
	
	 \textcolor{black}{Aerial FSO communications using an MRR present a viable solution for aircraft communications. This method, both cost- and energy-efficient, lightens the aircraft payload by eliminating the need for communication and beacon lasers and substituting them with an MRR~\cite{boeing90}. When the uplink beam is reflected in the MRR installed on the aircraft, the reflected beam automatically aligns with the direction of the gateway. As a result, the downlink pointing direction is unaffected by the posture instability and jitter of the aircraft. However, this method only allows for unidirectional communication since the payload of the aircraft can only modulate and reflect the incoming beam. Furthermore, a significant power loss occurs during the round-trip propagation of the signal.}

	\subsubsection{Liquid-Crystal}
	
	Liquid-crystal based beam-steering methods utilize an arrangement of a transmissive liquid-crystal layer to modulate the beam direction. This non-mechanical approach offers high precision control and low cost, making it an attractive enabler for PAT systems~\cite{2014advancedfso}. However, a key drawback is the limited dynamic angle range of liquid-crystal based modulators, restricted to a few milliradians. It presents challenges in supporting the rapidly changing tracking environment of aircraft communications. Additionally, liquid crystals have a relatively slow response time compared to electromechanical devices, which results in inadequate compensation for wide-bandwidth pointing disturbances of aircraft.

	\subsubsection{Beaconless PAT}

	 \textcolor{black}{In environments with reliably precise initial positioning and rare link switching, the beaconless PAT system is often the preferred choice. For satellite missions, the Consultative Committee for Space Data Systems (CCSDS) advocates the beaconless PAT approach for energy conservation~\cite{ccsds}. Utilizing appropriate detectors and actuators in PAT systems that depend exclusively on communication beams can significantly reduce the weight and operational complexity of the aircraft payload by eliminating the beacon transceivers. While this approach offers substantial cost benefits, the use of a beacon beam with an increased beam divergence is appealing for bidirectional aerial communications, especially when there is high angular movement and dynamic link scheduling.}

\begin{figure*}[t]
	\begin{center}
		{\includegraphics[width=1.15\columnwidth,keepaspectratio,frame]
			{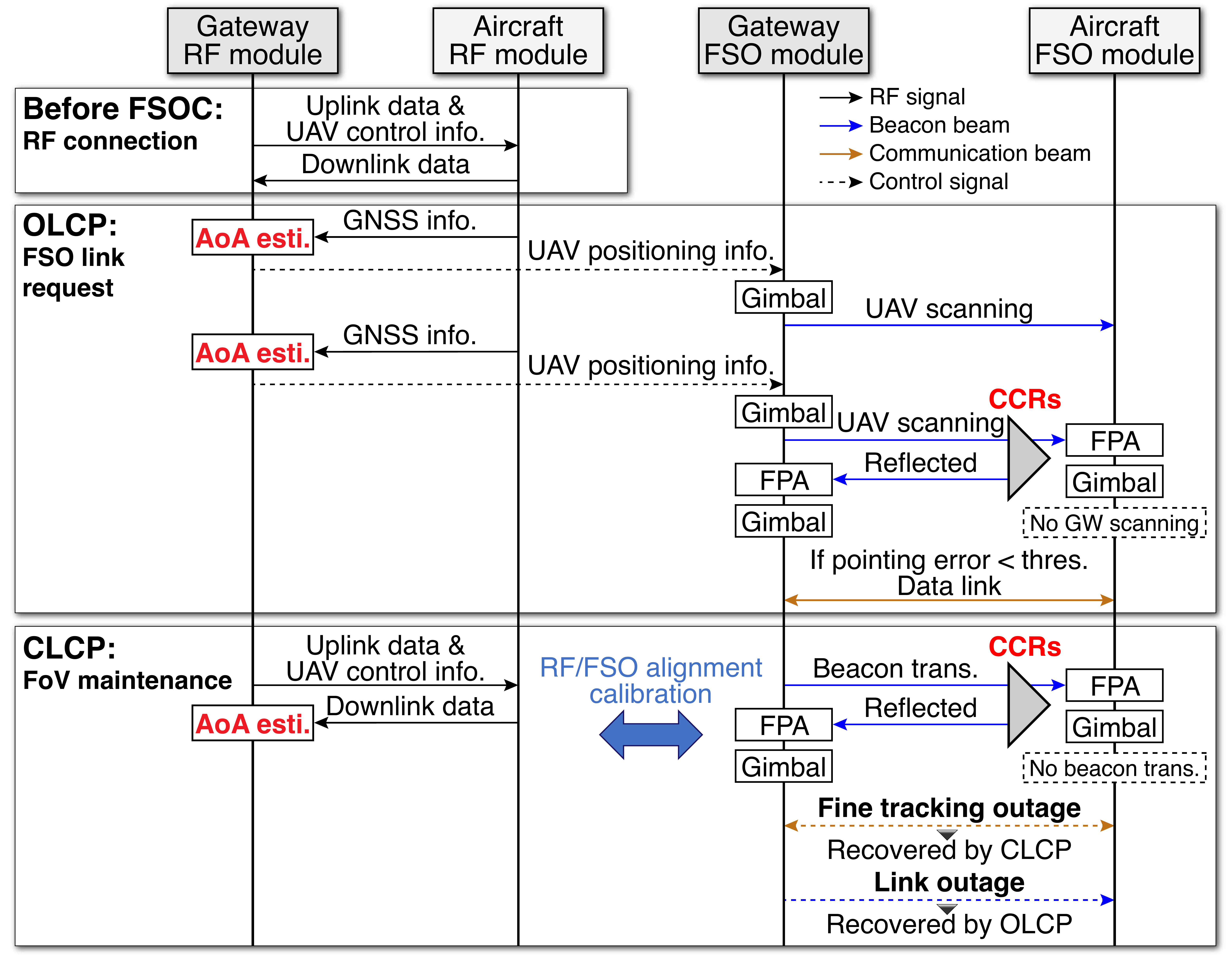}%
			\caption{The proposed pointing-and-acquisition algorithms for bidirectional FSO communications between the gateway and aircraft are depicted.}
			\label{fig03}
		}
	\end{center}
	\vspace{-15pt}
\end{figure*}

	\subsection{Positioning}
	
	During the OLCP process, a gateway requires precise positioning information of the aircraft. Therefore, it is generally assumed that the global navigation satellite system (GNSS) positioning information of the aircraft is first delivered to the gateway~\cite{boeing90}. When the gateway initially transmits the beacon beam to illuminate the aircraft, the scanning area depends on the primary positioning information error and the open-loop pointing error of the gimbal. For rapid backhaul link scheduling and acquisition among multiple ground gateways and aircraft, having precise positioning information for all possible aerial nodes is desirable at potential supporting gateways, whether managed in a centralized or distributed process.

	\subsection{Mobility}

	
	The pointing error caused by the movement of the aircraft is referred to as the point-ahead angle (PAA). It is widely considered to compensate for the PAA through biased control of the transmit beam using a point-ahead mirror in satellite communications~\cite{200710poi}. In the aerial FSO communication systems, the PAA can be effectively disregarded since the aircrafts travel at much lower speeds than satellites. A factor that is more crucial but difficult to analyze is the angular fluctuation resulting from posture changes in the aircraft during flight. It becomes especially critical when the FoV of the optical system and the dynamic range of the tracking actuators are limited. In such situations, a robust CLCP process is essential for maintaining the communication beam within the FoV of the detector. Considering the severe changes in aircraft pointing direction, in Section~\ref{sec4}, we propose and evaluate the pointing-and-acquisition algorithms that improve the outage performance and link-acquisition speed.

	\subsection{Others}

	
	\subsubsection{During OLCP}

	During the OLCP process, the gateway controls the gimbal and transmits a beacon beam to scan the aircraft. At this stage, open-loop gimbal control introduces pointing errors due to calibration error, step size, mechanical jitter, and thermal deformation~\cite{2020ssc}. 
	\textcolor{black}{Factors related to the posture instability of the aircraft can be avoided during the OLCP process if the gateway initiates the beacon transmission and the receiver FoV of the aircraft is sufficiently broad.}

	\subsubsection{During CLCP}

	During the CLCP process, factors outside the closed-loop control no longer affect the PAT performance. Instead, the feedback accuracy of the FPA measurement results in pointing errors. To be specific, as the payload estimates the incident direction of the received beam, noisy reception at the FPA creates an estimation error called noise equivalent angle (NEA). Also, the closed-loop pointing error occurs when the gimbal controller actuates the gimbal toward the estimated direction. \textcolor{black}{The body pointing error of the aircraft, determined by the accuracy of attitude sensing and control within the navigation system~\cite{201009jstqe}, also impacts the pointing error during the CLCP process. Furthermore, misalignment between the attitude sensor and the beacon transmitter contributes to pointing errors.}

	\subsubsection{During Fine Tracking}

	Due to the signal noise at the quadcell, beam spot detection by the quadrants generates an NEA during the fine-tracking process. The control signal is calculated linearly from the output voltage of the quadcell, as illustrated in Fig.~\ref{fig02}. However, the relationship between the actual incident direction and the output voltage is nonlinear, leading to a pointing error~\cite{201801access}. In addition, calibration residuals and control errors of the FSM, misalignment between the communication detector and the quadcell, and residual mechanical disturbances of the fine-tracking loop all contribute to the pointing error.

\begin{figure*}[t]
	\begin{center}
		{\includegraphics[width=1.15\columnwidth,keepaspectratio,frame]
			{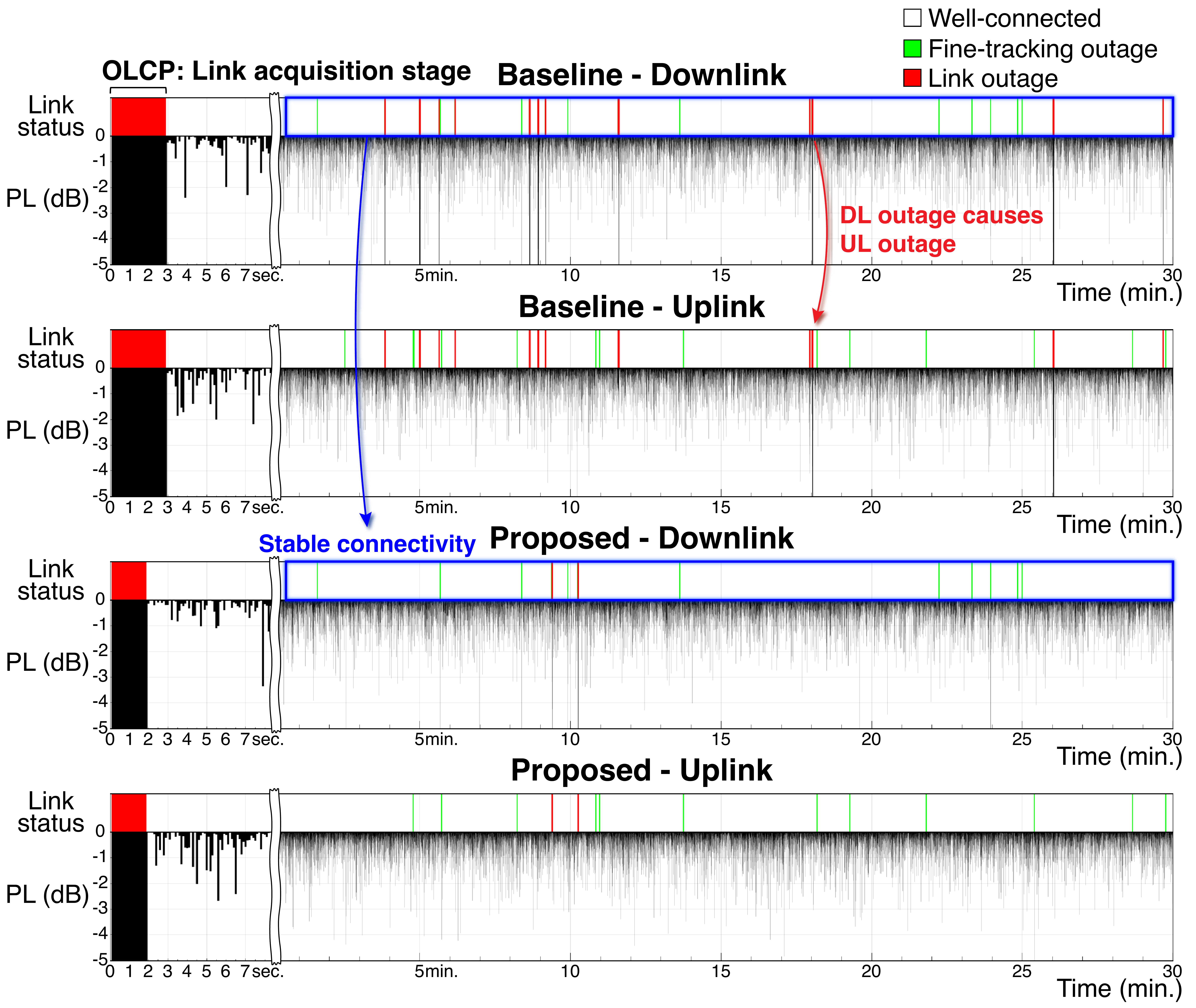}%
			\caption{A real-time link status and pointing loss (PL) of the baseline and proposed algorithms are presented. The proposed method shows shorter link-acquisition time and enhanced outage performance compared to the baseline method.}
			\label{fig04}
		}
	\end{center}
	\vspace{-15pt}
\end{figure*}


	\section{Pointing-and-Acquisition Algorithms and Performance Evaluation}
	\label{sec4}

	\subsection{Proposed Pointing-and-Acquisition Algorithms}
	
	\textcolor{black}{We propose novel pointing-and-acquisition algorithms for bidirectional FSO communications between a gateway and an aircraft, as illustrated in Fig.~\ref{fig03}.} The gateway RF module is equipped with either a planar or lens antenna array, allowing for the angle-of-arrival (AoA) estimation using the received RF signals~\cite{202202wcl}. Moreover, in our proposed system, beacon laser at the aircraft is replaced with multiple passive corner-cube reflectors (CCRs), reflecting the uplink beacon beam back to the gateway~\cite{2023eatvt}. In this system model, the CCRs selectively reflect the beacon beam using chromatic filters to avoid interference between the downlink communication beam and the reflected uplink communication beam. \textcolor{black}{In the following paragraphs, we will describe the operation scheme of the ground and aerial payloads before the FSO communications, during the OLCP, and during the CLCP processes.}
	

	\subsubsection{Before FSO Communications}
	
	Network management systems and aircraft decide whether to establish an FSO link between a particular gateway and aircraft based on link availability between the nodes and an efficient routing path for services. Especially when a gateway can serve a limited number of aircraft simultaneously, a link scheduling process by the network management system and rapid link acquisition are necessary. Aircraft can then be flexibly deployed as mobile base stations of the 6G network, supporting high data rates and massive connectivity in unserved, disaster, and temporarily crowded areas. For these applications, we assume that the aircraft continuously exchange control and user data with ground gateways via RF links. The network management system can request FSO connections immediately in this circumstance.

	\subsubsection{OLCP for Link-acquisition}
	
	When an FSO link is requested between an aircraft and a gateway, the aircraft transmits its GNSS information via the RF link. In the proposed algorithm, the gateway receives the GNSS information and estimates the AoA of the signal using the antenna array. \textcolor{black}{By integrating the GNSS information with the estimated AoA via the maximum likelihood criterion, the gateway controls the gimbal~\cite{202202wcl}.} It then transmits a beacon laser to scan the aircraft until the aircraft receives the beacon beam through the FPA and aligns the gimbal pointing. In the baseline algorithm, a beacon laser at the aircraft transmits a beacon beam back to the gateway to achieve an alignment of the gateway payload using the FPA detection of the beam. However, in the proposed method, the CCRs deployed around the payload reflect the uplink beacon beam back to the gateway. In other words, the payloads are aligned using FPA detection like the baseline algorithm, but CCRs replace the beacon laser at the aircraft. \textcolor{black}{When the direction of the incident beam falls within the FoV of the communication detector, they initiate communication and start the fine-tracking and CLCP processes.}
	
	\subsubsection{CLCP for Link-maintenance}
	
	The CLCP process compensates for large angular movements of entire payloads using gimbals. The gateway transmits a beacon beam and captures the reflected beam, while the aircraft only receives the beam through the FPA. Both sides detect the misalignment and utilize closed-loop gimbal control to keep the received beam spot centered on the FPA. This process also protects the maximum dynamic range of the FSM by initializing the tilted angle of the FSM. When an outage occurs, the terminals can reconstruct the link through the OLCP process. For precise coordination between the RF module and gimbal controller when reconstructing the link, AoA estimation is continuously performed during the FSO communications to ensure accurate mapping between the RF AoA and gimbal control.

	\textcolor{black}{\subsection{Advantages of the Proposed Algorithms}}
	
	\textcolor{black}{As depicted in Fig.~\ref{fig03}, our proposed algorithm introduces two methods. The first combines AoA estimation of downlink RF signals at the gateway with GNSS data, and the second involves passive CCRs placed around the FSO transceiver at the aircraft. These enhancements augment both the stability and the link-acquisition speed, as verified by our simulations.}
	
	\subsubsection{RF AoA Estimation}

	\textcolor{black}{When GNSS information of the aircraft is acquired by the gateway to point the aircraft during the initial link acquisition, the RF AoA estimation improves the pointing accuracy. The study in~\cite{202202wcl} reveals that a simple linear combination of GNSS data and AoA estimation can reduce outage probability by factors ranging from multiple to hundreds, depending on channel conditions. In our algorithms, we also utilize this approach to restore the link during unexpected outages by continuously implementing the AoA estimation during FSO communications.}

	\subsubsection{Deployment of the CCRs at the Aircraft}
	
	\textcolor{black}{Deploying multiple passive CCRs around the aircraft transceiver offers numerous benefits. Given that a beacon link does not require high-frequency modulation, uplink beacon signals can be repurposed for downlink via low-cost passive CCRs. Therefore, it can replace the role of the beacon laser in the aircraft. From an energy-saving perspective, this can achieve the same energy consumption at the aircraft as the recommended beaconless system~\cite{ccsds}. In addition, the retroreflective nature of the CCR eliminates the need for downlink beacon beam pointing. In the process of the uplink beacon beam reflecting down, the diverse beams from these CCRs allow the beacon receiver at the gateway to acquire a diversity effect. The results presented in~\cite{2023eatvt} show that this diversity yields great advantages against outages.}

	\textcolor{black}{\subsection{Performance Metrics}}

	\textcolor{black}{We categorize outages into two types to evaluate the proposed algorithms from an outage perspective.}
	
	\subsubsection{Fine-tracking Outage}
	
	\textcolor{black}{This outage occurs when the misalignment surpasses the FoV of the quadcell, or when the signal power the quadcell receives is too weak due to deep fading or a sudden surge in pointing errors. When the outage occurs, the pointing loss of the communication beam increases as the jitter cannot be compensated by the fine-tracking system. However, the CLCP process can quickly restore the fine-tracking system by adjusting the gimbal pointing.}
	
	\subsubsection{Link Outage}

	\textcolor{black}{The link outage occurs when the beacon beam is no longer detected at the FPA due to power outages or drastic changes in payload pointing direction. In such cases, the system initiates the OLCP process to reestablish the link. If communication is unavailable due to weather conditions, the two terminals must rely on RF links or have the network management system route the data through other links.}
	

	\textcolor{black}{\subsection{Simulation Results}}

	\textcolor{black}{During the simulation in MATLAB, we assume that the coherence time of the random channel is $0.1$~s. For each channel slot, we operate the detectors and tracking actuators of our pointing-and-acquisition systems. We define five states: link request, OLCP process, well-connected, fine-tracking outage, and link-outage. Success in the OLCP process or an outage during the well-connected state leads to state changes for individual links, such as the uplink beacon, downlink beacon, uplink communication, and downlink communication links.}

	We set the link distance to $2$~km and the standard deviation of GNSS error to $5$~m~\cite{202202wcl}. The parameters related to the PAT devices, the quadcell FoV, FPA FoV, and CLCP loop frequency, are set to $2$~mrad, $40$~mrad, and $1$~Hz, respectively. We assume the standard deviation of the residual errors for the open-loop gimbal control, closed-loop gimbal control, and FSM control to be $3$~mrad, $0.3$~mrad, and \SI{100}{\micro\radian}, respectively. The visibility range of the FSO channel is $3$~km, and the gamma-gamma fading channel with strong turbulence is assumed~\cite{2014advancedfso}. The communication and beacon beam divergence angles are~\SI{500}{\micro\radian} and $5$~mrad, respectively. \textcolor{black}{In the retroreflective channel, reciprocity is considered by sampling the correlated uplink and downlink random channels with correlation coefficients of $0.4$ and $0.7$. The chosen correlation coefficients are based on wave optical calculations~\cite{201205ao}.} The positioning estimation for the aircraft is performed using the GNSS information and AoA-estimation results as proposed in~\cite{202202wcl}, and 4 CCRs are circularly deployed around the aircraft payload.

	The simulation results indicate that the proposed algorithm offers a reduced outage probability and a shorter average link-acquisition time. Fig.~\ref{fig04} depicts the link status and pointing loss over a 30-minute mission flight. The mission begins in the link-outage state, with the OLCP process attempting to establish a connection. During the communication phase, the link outage mainly occur due to the posture instability of the aircraft. Therefore, the fixed beam pointing of the reflected beam and the spatial diversity effect achieved by multiple CCRs significantly contribute to the stability of the FSO link.
	
\begin{figure}[!t]
	\begin{center}
		{\includegraphics[width=0.8\columnwidth,keepaspectratio,frame]
			{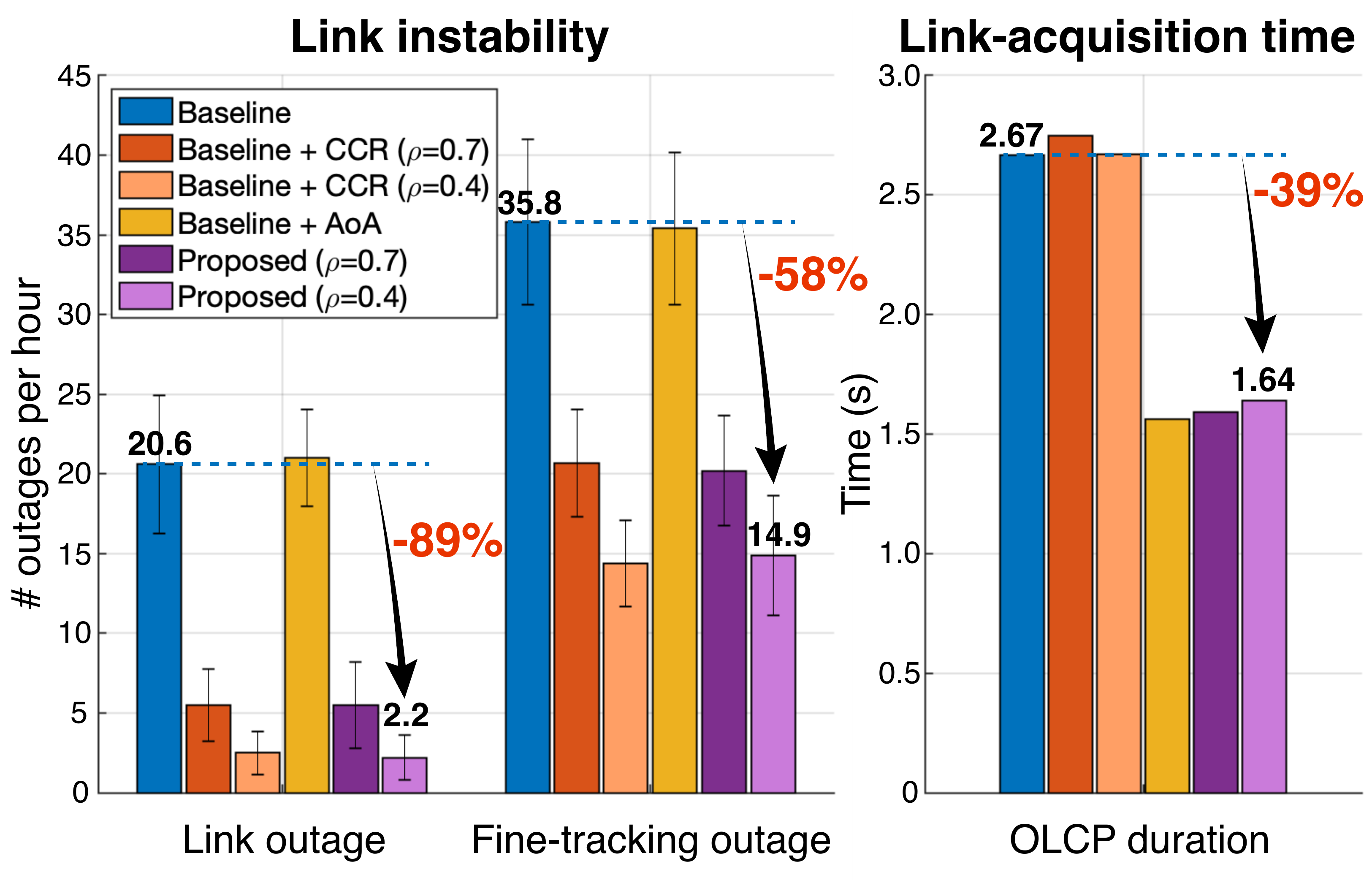}%
			\caption{\textcolor{black}{Performance evaluation for pointing-and-acquisition simulations. Note that 'Baseline + CCR' and 'Baseline + AoA' are also novel techniques simulated for comparison.}}
			\label{fig05}
		}
	\end{center}
	\vspace{-15pt}
\end{figure}

	\textcolor{black}{In Fig.~\ref{fig05}, we contrast the performance of the proposed algorithm against the baseline algorithm, the baseline algorithm augmented with the AoA estimation technique, and the baseline algorithm enhanced with CCR deployment. The results show that the average number of link outages over a one-hour flight is considerably reduced for both cases, where the correlation coefficient of the retroreflective channel is set to $0.4$ and $0.7$, respectively. Moreover, when the AoA-estimation technique is incorporated, the average link-acquisition time significantly decreases due to enhanced positioning accuracy during the OLCP process.}
	
\begin{figure}[!t]
	\begin{center}
		{\includegraphics[width=0.8\columnwidth,keepaspectratio,frame]
			{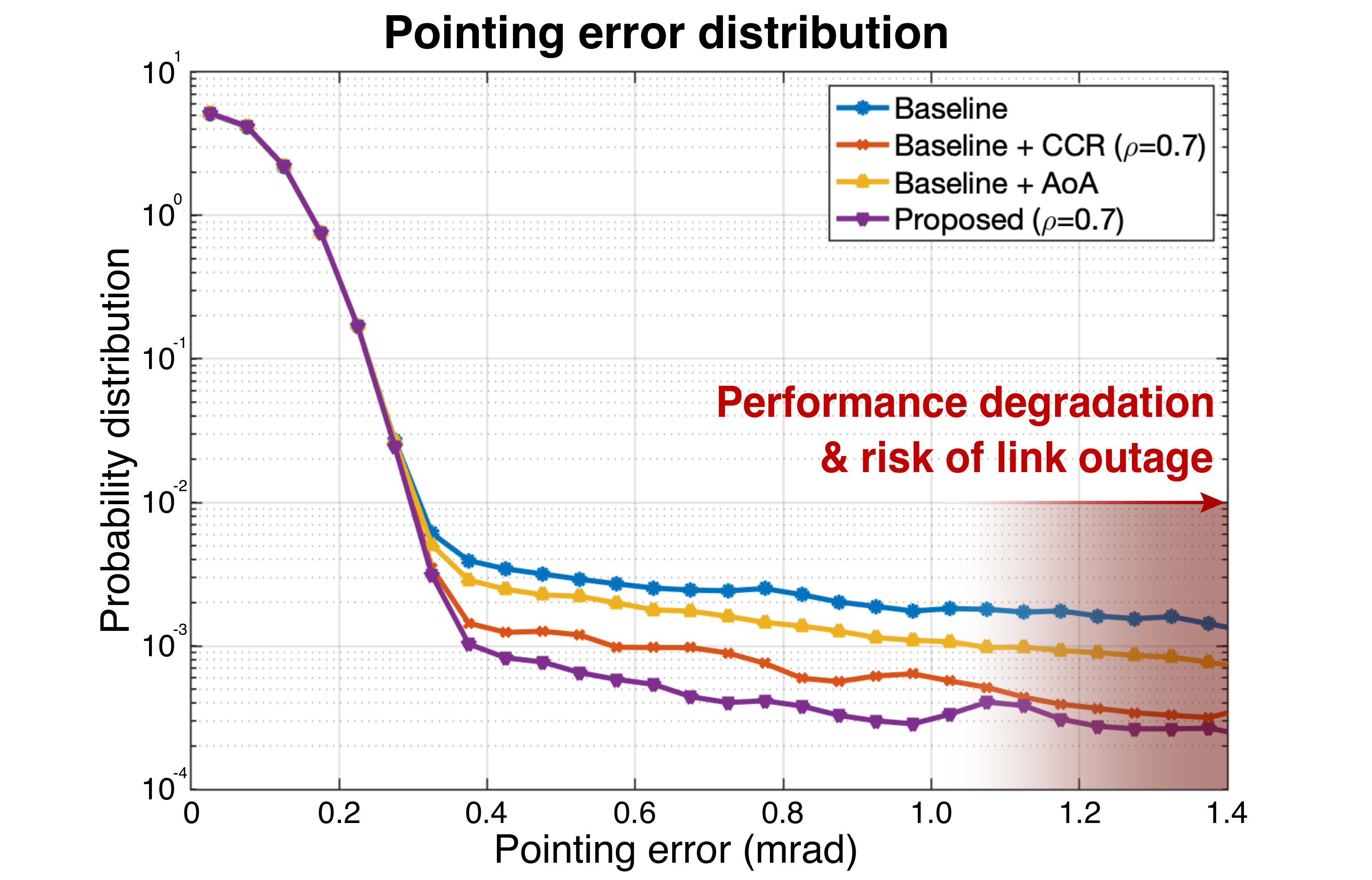}%
			\caption{\textcolor{black}{Distribution of the simulated downlink pointing error.}}
			\label{fig06}
		}
	\end{center}
	\vspace{-15pt}
\end{figure}

	\textcolor{black}{Fig.~\ref{fig06} represents the pointing error distribution of the downlink communication beam during the simulation. When the AoA-estimation method is utilized, the link can quickly recover from a large pointing error. The deployment of CCRs on the aircraft enhances the robustness of the beacon tracking system, thereby minimizing the pointing error. Lastly, the results for our proposed algorithm show that we can significantly suppress the pointing error by combining both techniques.}

	\section{Conclusion}
	{In this article, we highlighted the challenges of vertical FSO communications from the perspective of a PAT. The bidirectional connectivity of vertical links is influenced by various factors, including payload constraints, positioning accuracy, mobility, and other physical limitations of aircraft. Based on the comprehensive investigation, we developed the baseline PAT algorithms and novel pointing-and-acquisition algorithms. The simulation results showed the enhanced performance of the proposed algorithms using AoA estimation and retroreflectors. In conclusion, this article provides valuable insights into aerial FSO communications as an enabler of the future integrated ground-air 6G networks.}

\bibliographystyle{IEEEtran}
\bibliography{Moon_mag_pat} 	


\textbf{Hyung-Joo Moon} (Graduate Student Member, IEEE) is currently pursuing a Ph.D degree in the School of Integrated Technology, Yonsei University, South Korea.

\textbf{Chan-Byoung Chae} (Fellow, IEEE) is an Underwood Distinguished Professor at Yonsei University, South Korea. His research interest includes emerging technologies for 6G.

\textbf{Kai-Kit Wong} (Fellow, IEEE) is with the Dept. of Electronic and Electrical Engineering, University College London, UK. He is EiC for the IEEE Wireless Comm. Letters. 

\textbf{Mohamed-Slim Alouini} (Fellow, IEEE) is a Distinguished Professor of Electrical Engineering at King Abdullah University of Science and Technology (KAUST), Saudi Arabia. His current research interests include the modeling, design, and performance analysis of wireless communication systems.

\end{document}